\documentclass[prb,showpacs,preprintnumbers,twocolumn]{revtex4}

\usepackage{amssymb}
\usepackage{color}
\usepackage{graphicx}
\def\be{\begin{equation}}
\def\ee{\end{equation}}
\def\bea{\begin{eqnarray}}
\def\eea{\end{eqnarray}}

\begin{document}

\title{Universality of a family of Random Matrix Ensembles with logarithmic soft-confinement potentials}
\author{Jinmyung Choi and K.A. Muttalib}
\affiliation{Department of Physics, University of Florida, Gainesville FL 32611-8440}
\keywords{}
\pacs{72.15.Rn, 05.40.-a, 05.60.-k, 05.90.+m, 71.30.h}

\begin{abstract}
Recently we introduced a family of $U(N)$ invariant Random Matrix Ensembles which is characterized by a parameter $\lambda$ describing logarithmic soft-confinement potentials $V(H) \sim [\ln H]^{(1+\lambda)} \:(\lambda>0$). We showed that we can study eigenvalue correlations of these ``$\lambda$-ensembles" based on the numerical construction of the corresponding orthogonal polynomials with respect to the weight function $\exp[- (\ln x)^{1+\lambda}]$. In this work, we expand our previous work and show that: i) the eigenvalue density is given by a power-law of the form $\rho(x) \propto [\ln x]^{\lambda-1}/x$ and ii) the two-level kernel has an anomalous  structure, which is characteristic of the critical ensembles. We further show that the anomalous part, or the so-called ``ghost-correlation peak", is controlled by the parameter $\lambda$; decreasing $\lambda$ increases the anomaly.  We also identify the two-level kernel of the $\lambda$-ensembles in the semiclassical regime, which can be written in a sinh-kernel form with more general argument that reduces to that of the critical ensembles for $\lambda=1$.  Finally, we discuss the universality of the $\lambda$-ensembles, which includes Wigner-Dyson universality ($\lambda \to \infty$ limit), the uncorrelated Poisson-like behavior ($\lambda \to 0$ limit), and a critical behavior for all the intermediate $\lambda$ ($0<\lambda<\infty$) in the semiclassical regime. We also comment on the implications of our results in the context of the localization-delocalization problems as well as the $N$ dependence of the two-level kernel of the fat-tail random matrices. 
 \end{abstract}

\maketitle


\section{Introduction}
Random matrix theory (RMT) deals with the statistical properties of eigenvalues and eigenvectors of random matrices drawn from a certain probability measure. The theory was  successfully applied by Wigner \cite{Wigner} in the 1950s to describe the spectral properties of complex many-body nuclei where the underlying Hamiltonians are so complicated  that a useful way to study the system turned out to be through a statistical treatment of the Hamiltonians \cite{Dyson}. Since then, RMT has been applied to a wide variety of systems in diverse areas, including e.g.  many-body atoms and nuclei, quantum chaos, mesoscopic disordered conductors, 2-D quantum gravity, conformal field theory, chiral phase transitions as well as zeros of Riemann zeta function, scale-free networks, biological networks, communication systems and financial markets \cite{reviews, rmt-complex}.
This broad range of applicability of RMT in seemingly unrelated areas highlights the universal features of the correlations of the eigenvalues in RMT. Within the classical Gaussian model pioneered by Wigner, these correlations are known as the Wigner-Dyson (WD) statistics of the Gaussian ensembles, which are qualitatively different from the statistical features of completely uncorrelated eigenvalues given by the Poisson statistics. 

In the past decades, many attempts have been made to construct \textit{generalized} random matrix ensembles that incorporate \textit{power-law} or \textit{fat-tail} distributions. \cite{CB,burda,fat-tail-rmt}. The significance of such generalization beyond the Gaussian ensembles is mainly two fold. First of all, it is conceivable that the wide applicability of classical or Gaussian RM models is closely linked to the prevalence of normal or Gaussian distributions in nature, a consequence of the central limit theorem \cite{CB}. It has, however, not been fully investigated if there is a counterpart of the Gaussian ensemble as implied by the structure of the \textit{generalized} central limit theorem \cite{CB}, comparing e.g, the Gaussian and the L\'evy  basins. Second, there have been expanding interests in RMT applications to generic complex systems such as financial markets, scale-free network, earthquakes etc. \cite{rmt-complex,complex} that feature fat-tail noise, but the relevance of the classical RM models in these systems seems questionable. This is because the Gaussian ensemble is based on the assumption that the systems are characterized by Gaussian noise, which is clearly not suitable for systems with fat-tail noise where the occurrence of extreme events are not as rare as expected from normal distributions. So far, fat-tail distributions in random matrix ensembles have been carefully incorporated in some limited cases and the calculation of the correlation functions of eigenvalues have been carried out for certain special cases \cite{fat-tail-rmt}. However, the question regarding the universality of the correlations of the  eigenvalues remains unresolved. For Gaussian ensembles it is the well known two-level sine kernel that establishes the universality of the correlations in the properly scaled large $N$ (matrix size) limit; it is not clear if there exists a similar universal two-level kernel for the power-law or the fat-tail ensembles as well.

In fact, the suggestion of  a novel universality beyond the Gaussian ensembles comes from the study of the Anderson transition in the disordered electronic systems \cite{critical}. In these systems, the Gaussian ensemble is only relevant in the metallic regime where all the eigenstates are extended across the entire system and correlations of the corresponding eigenvalues are well described by the WD statistics.  As the disorder is made strong enough, the eigenstates become localized and thus the eigenvalues become uncorrelated. Especially at the delocalization-localization transition, it has been established that the correlations of the eigenvectors exhibit novel features \cite{critical} such as multi-fractality and the correlations of the eigenvalues lead to a level compressibility that is intermediate between WD and Poisson statistics. Similarly in the studies of quantum chaos, energy level statistics of  systems that are intermediate between chaotic and regular states also require generalization beyond WD and Poisson statistics \cite{crossover}. 
In these contexts, extensive studies have been carried out to construct a parametric generalization of RM models  that cross over from WD to Poisson \cite{critical,crossover} as a function of the parameter. Some of these generalizations indeed capture the essential features of the critical statistics, among which the family of q-RMEs \cite{qrme1,qrme2}  provides a particularly valuable insight. Within the common framework of rotationally invariant RM models \cite{Mehta} the q-RMEs show how the universality of the Gaussian ensemble characterized by the well-known \textit{sine} kernel breaks down and eventually gives rise to a different kind of universality for the critical ensembles, characterized by a one-parameter two-level \textit{sinh} kernel. In particular, the rotationally invariant RM models are characterized by a ``confining potential'' which defines the weight function of a set of orthogonal polynomials; the key difference between the Gaussian and the critical ensembles comes from the fact that the corresponding orthogonal polynomials, namely classical vs. $q$-orthogonal polynomials, respectively, possess qualitatively different asymptotic properties  \cite{szego,ismail}.

In the motivation to investigate the universality associated with fat-tail or power-law RMEs, we introduced a family of $U(N)$ invariant random matrix ensembles characterized by an asymptotic logarithmic potential $V(H) = A [\ln x ]^{1+\lambda}$ with $\lambda>0$ \cite{cm-jpa09}, which we will refer to as ``$\lambda$-ensembles"  \footnote{The earlier name `L\'evy like ensembles' in Ref.~[\onlinecite{cm-jpa09}] reflects the motivation of the study.}. The reason for such a choice of the potential is based on the following few observations.  First, it is known that for $V(H) \propto [\ln H]^2$ corresponding to $\lambda = 1$ (the critical ensemble), the eigenvalue spectrum is given by inverse power-law distribution, as known by the mean-field theoretic approach \cite{mean-field}. Second, for $V(H) \simeq N \ln H $  corresponding to the $\lambda \to 0$ limit with the constant $A$ being order of $N$ (free L\'evy matrices), the spectral density is given by the fat-tail distributions. Third, in the limit $\lambda \gg 1$, it is expected that the confinement potential may grow sufficiently strong, thereby approaching the Gaussian limit. The fact that such a parametric generalization connects various existing RM models is interesting since the model allows us to explore any possible novel universality associated with fat-tail RMEs and with logarithmic soft-confinement potentials within the rotationally invariant RMT framework.

It is well known from the mean-field theoretic approach \cite{mean-field} that RMEs with logarithmic soft-confinement are characteristically different from those with strong confinement potentials given by $V(x) = |x|^\alpha, \alpha>1$. The reason is that for the soft-confinment potential, the eigenvalue density does not depend on $N$ and has a non-trivial functional form, which is not translationally invariant. It means that the required unfolding is non-trivial.  In fact, the non-trivial unfolding procedure give rise to the deviation from the WD statistics. In the following, we will show by studying the two-level kernel that the $\lambda$-ensembles exhibits interesting deviations from WD statistics. Especially, the presence of the anomalous component in the two-level kernel is a common characteristics of all the logarithmic confinement potentials. 

The paper is organized as follows. In sec II we provide a brief review of the orthogonal polynomial method and in sec III we discuss the model ensemble. In sec IV we will show the results and finally in sec V we will discuss the results with concluding remarks.

\section{Orthogonal Polynomial Method}

We consider the set of $U(N)$ invariant Hermitian matrices $H$ with the following probability measure
\begin{equation}
P_N(H) dH \propto e^{-\textrm{tr}[V(H)]} dH,
\end{equation}
where the confining potential $V(x)$ is a suitably increasing function of $x$, 
$\textrm{tr}$ is the matrix trace and $dH$ the Haar measure. 
In the eigenvector basis, the joint probability distribution of the
eigenvalues $\{x_i\} (i=1,2,\dots N)$ of the matrices can be written in the form \cite{Mehta}
\begin{equation}
P_N(\{x_i\})\propto 
\prod_{ i<j }^N (x_i - x_j)^2 \prod_{i=1}^N e^{-V(x_i)}.
\end{equation}
Here the factor $\prod (x_i-x_j)$ is the Vandermonde determinant. 
Given a set of (monic) polynomials $p_n(x)$ that 
are orthogonal with respect to the weight function $w(x)=e^{-V(x)}$, i.e.  
\begin{equation}
\int\limits_{-\infty}^{\infty} e^{-V(x)} p_n(x) p_m(x)dx
=\delta_{mn},
\end{equation}
the n-level correlation of eigenvalues $\mathcal{R}_n (x_1,x_2, ...., x_n)$ can be written in a compact form in terms of the
the two-level kernel in the following manner,
\begin{equation}
\mathcal{R}_n (x_1,x_2, ...., x_n)\equiv \det[K_N(x_i,x_j)]_{ \{i,j=0,...,n\} }. 
\end{equation}
Here the two-level kernel $K_N(x,y)$ is defined as  
\begin{equation}
K_N(x,y)\equiv \sum_{n=0}^{N-1} \psi_n(x) \psi_n(y), 
\label{K-N} 
\end{equation}
where the ``wave function'' $ \psi_n(x) \equiv p_n(x)e^{-V(x)/2}$.  By using the Christoffel-Darboux formula, the kernel is simplified to (for the monic polynomials)
\begin{equation}
\label{C-D}
K_N(x,y) =  \frac{\psi_N (x) \psi_{N-1}(y)- \psi_N(y)\psi_{N-1}(x)}{x-y} . 
\end{equation}
In general, the large $N$ asymptotic behavior of the orthogonal polynomials with respect to any weight function characterized by $V(x) \sim x^{\alpha}$, $\alpha > 1$, have behavior qualitatively similar to the Hermite polynomials \cite{qrme2}, such that the asymptotic behavior of the wave function in the $N \to \infty$ limit is given by
\begin{equation} 
\psi_{2N}(u) \sim \cos(\pi u) ~~;  \psi_{2N-1}(v) \sim \sin(\pi v) 
\end{equation}
where  $u$ and $v$ are scaling variables, namely, d$u \equiv K_{\infty}(x,x) dx$ and $dv \equiv K_{\infty}(y,y) dy$ such that the mean density $\bar{K}_{\infty}(u,u)$ is unity in the N $\to \infty$ limit. In this limit, the two-level kernel in the scaled variables become
\begin{equation}
\label{K-G} 
\bar{K}^G(s)=\frac{sin(\pi s)}{\pi s}; \;\;\; s \equiv (u-v) 
\end{equation}
which is the celebrated sine kernel.
Thus, in the framework of orthogonal polynomial method, the universality of random matrix ensembles is traced to the similarity of the scaling behavior in the large $N$ limit of the polynomials corresponding to the probability measure under consideration \cite{qrme2}.  
In particular, the orthogonal polynomials corresponding to all `Freud-type' weight functions  $e^{-V(x)}$, with monotonically increasing polynomial $V(x)$, share similar asymptotic behavior in the large $N$ limit. Thus, they give rise to the same correlations as the Gaussian ensembles. 

\section{The $\lambda$-ensembles} 
The generic choice of the confining potential $V(x)$ that gives asymptotic logarithmic behavior is $V(x) = A [\ln x] ^{1+\lambda}$. However, it has an unphysical singularity at the origin so that we need to regularize it in a certain way. One possible way to do it is choosing, \textit{e.g.}, $V(x) = A [\ln (1+x)] ^{1+\lambda}$ but there are a variety of other forms that differ by the regularization behavior in the vicinity of origin, which will not change the characteristics of the $\lambda$-ensembles. For our study, we particularly choose the following form of potential \cite{cm-jpa09}:
\begin{eqnarray} \label{eq-Vx}
V (x) &=& \frac{1}{\gamma}[\sinh^{-1}x]^{1+\lambda}; \;\;\; \lambda > 0, \;\;\; \gamma>0
\end{eqnarray}
The merit of choosing Eq.~(\ref{eq-Vx}) is that for $\lambda = 1$, it coincides with the one possible form of the weight function of the $q$-RMEs \footnote{In which case, $\lambda=1$ and $\gamma = \ln(1/q) \; q>1$} so that we can compare our results with those of $q$-RMEs. For the $q$-RMEs, the mathematical properties of the corresponding orthogonal polynomials, ``the Ismail-Masson $q$-polynomials" \cite{ismail} are well known, which leads to the two-level $\sinh$ kernel in the limit $\gamma \ll 2\pi$, given by
\begin{equation}
\label{K-C}
\bar{K}^C(s) = \frac{\gamma}{2\pi} \frac{\sin(\pi s)}{\sinh (\gamma s/2)}.
\end{equation}
As is well known, this recovers the Gaussian sine kernel in the limit $\gamma =0$.

For $\lambda \neq 1$, the corresponding orthogonal polynomials, which we refer to as ``$\lambda$-generalization of $q$-polynomials" \footnote{Note that In ref. ~[\onlinecite{cm-jpa09}], it is introduced as ``generalized $q$-polynomials".  Here we use ``$\lambda$-generalization of $q$-polynomials"  in order to clarify the meaning of ``generalization" and to emphasize the role of the parameter $\lambda$.} 
are known via the recurrence relation of the orthogonal polynomial, \textit{i.e.}, 
\begin{equation}
x\phi_n(x)=\phi_{n+1}(x)+S_n\phi_n(x)+R_n\phi_{n-1}(x),
\end{equation}
where $S_n$ and $R_n$ are real. For symmetric weight functions all $S_n$ are zero, and the polynomials are determined by the coefficient $R_n$. The following shows comparison among the Hermite polynomials, $q$-polynomials, and the $\lambda$-generalization of $q$-polynomials:
\begin{eqnarray}
R_n &\propto & n \;\;\;\; \textrm{Hermite polynomials,}\\
	&\propto & e^{\gamma n} \;\;\;\; \textrm{q-polynomials,} \\
	&\propto & e^{\gamma n^{1/\lambda}} \;\; \textrm{$\lambda$-generalization of $q$-polynomials.} \label{Rn-lambda}
\end{eqnarray}
We can see clear distinction between $R_n$ of the Hermite polynomials (linear in $n$) and that of the $\lambda$-generalization of $q$-polynomials (exponential in $n^{1/\lambda}$) which reduces to that of $q$-polynomials for $\lambda =1$. The significance of $R_n$ is that it determines the upper bound of the spectral density and thus the scaling behavior of the bulk of the spectrum in the large $N$ limit \cite{szego,ismail}. For example, for the Gaussian ensembles characterized by $V(x) = x^2$ (Hermite polynomials), the upper bound of the spectral density is $\sqrt{N}$ in the large $N$ limit. Thus the normalization condition of the spectral density requires the bulk of the spectrum to grow as $\sqrt{N}$ as in the semi-circle law. On the other hand for the logarithmic soft-confinement potentials, the spectral edge grows at an exponential rate $e^{\gamma n^{1/\lambda}}$. The bulk of the spectrum does not scale as $N$. For example, for $V(x) \propto [\ln x]^2$, the spectral edge grows as $\sqrt{e^N}$ and the bulk of the spectrum is given by $\rho(x) \propto 1/x$ which does not depend on $N$.

\section{Results}
The calculations of the density and the cluster function are performed based on the numerical construction \cite{spm} of the $\lambda$-generalization of the $q$-polynomials. For instance, the density $\rho(x) \equiv K_N(x,x)$ can be obtained by summations of products of the wave function as in Eq.~(\ref{K-N}). The two-level cluster function 
\begin{equation}
Y(u,v) \equiv [\bar{K}(u,v)]^{2} 
\end{equation}
can be obtained based on Eq.~(\ref{K-N}) with the unfolding map $u(x) \equiv \int^{x} dx \rho (x)$.

\subsection{Eigenvalue Density}

In Ref.~[\onlinecite{cm-jpa09}] it was observed that the density of the $\lambda$-ensembles is given by a pure power-law, e.g., $ \rho(x) = \frac{1}{x^{1-\theta}}$. For $\lambda=1$, $\theta =0$ and for $\lambda > 1$ and $\lambda <1$, $\theta >0$ and $\theta<0$, respectively. However, careful further investigation shows that this observation is only approximate and a more accurate form of the eigenvalue density is given by
\begin{equation}
\label{eq-density}
\rho(x) \propto \frac{[\ln x]^{\lambda-1}}{x} \;\;\; \textrm{for x~} \gg \Lambda
\end{equation}
where the lower cutoff $\Lambda$ depends on the parameters in $V(x)$. For example for $\lambda=1$, $\Lambda= \gamma$ and for $\lambda \neq 1$, $\Lambda$ is some function of both $\gamma$ and $\lambda$, which depends on the specific regularization of $V(x)$ at the origin.

\begin{figure}[tbp]
\begin{center}
\includegraphics[angle=0, width=0.32\textheight]{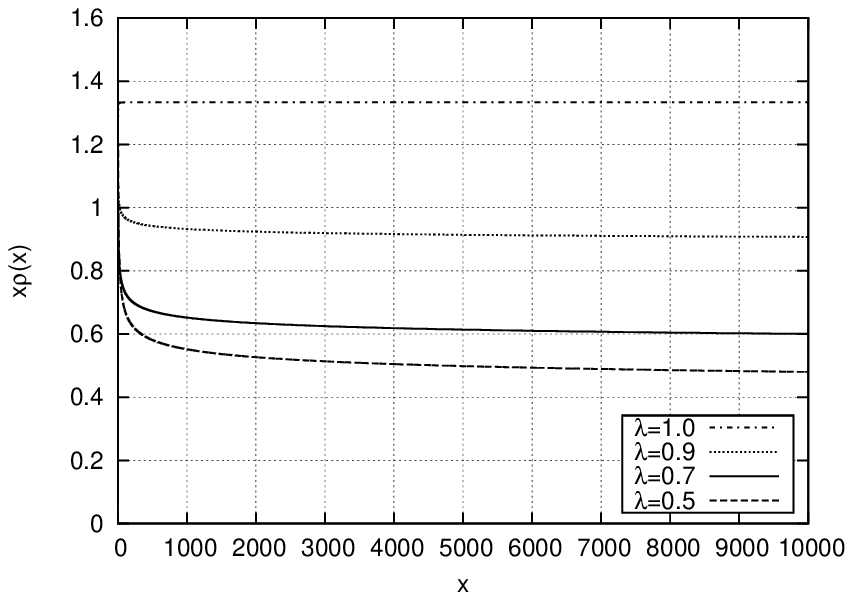}
\end{center}
\caption{Eigenvalue density for $\lambda \le 1$, namely $\lambda = 0.9$ ($\gamma$ = 0.75), 0.7 ($\gamma$ = 0.50), 0.5 ($\gamma$ = 0.25) as well as 1 ($\gamma$ = 0.75).} 
\label{KNxx-01}
\end{figure}


\begin{figure}[tbp]
\begin{center}
\includegraphics[angle=0, width=0.32\textheight]{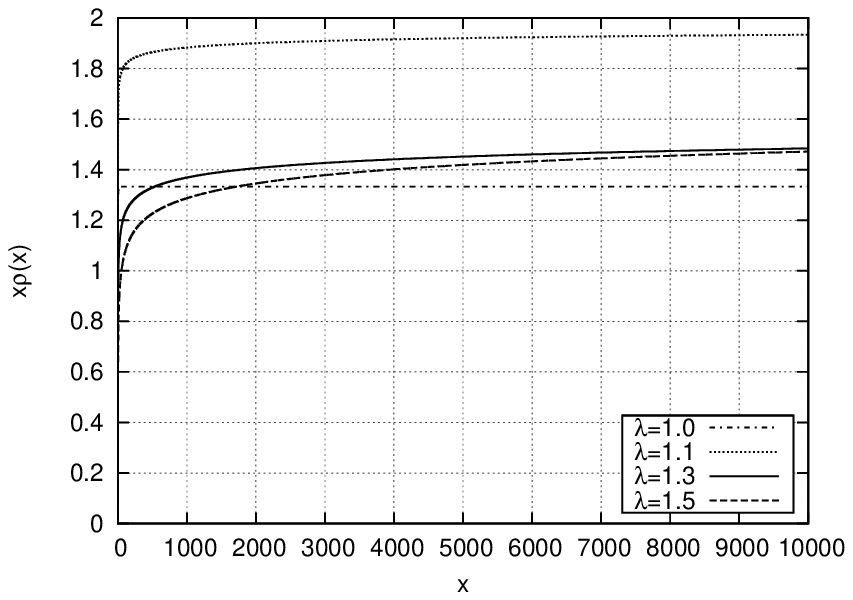}
\end{center}
\caption{Eigenvalue density for $\lambda \ge 1$, namely $\lambda = 1.1$ ($\gamma$ = 0.75), 1.3 ($\gamma$ = 2.00), 1.5 ($\gamma$ = 4.00) as well as 1 ($\gamma$ = 0.75).} 
\label{KNxx-03}
\end{figure}

The above form is suggested by the mean-field approach \cite{Kravtsov} for $\lambda \geq 1$. The validity of this form for all $\lambda>0$ can be checked by considering the normalization condition of the spectral density 
\be
2\int_{0}^{D_N} \rho (x) dx = N.
\ee 
Here the factor 2 comes from the fact that $\rho(x)$ is symmetric around origin. The upper bound $D_N$ is given by the largest zero of the orthogonal polynomials of order $N$, namely $D_N \propto \sqrt{R_N}$. As pointed out, $R_n \propto \exp [n^{1/\lambda}]$. 
We notice that Ref. [\onlinecite{chen-ismail}] studied the largest zeros of the orthogonal polynomials to the weight function of $\exp[-c (\ln x)^m ]$ for $c>0$ and $m$ a positive even integer and showed that the largest zero is of order $\exp( n^{\frac{1}{m-1}})$, which is the same behavior as the coefficient $R_n$ of the $\lambda$-generalization of $q$-polyonomials in Eq.~(\ref{Rn-lambda}). Thus, our results seem to imply that the results of Ref. [\onlinecite{chen-ismail}] can be extended to an arbitrary real $\lambda >0$.

\begin{figure}[tbp]
\begin{center}
\includegraphics[angle=0, width=0.32\textheight]{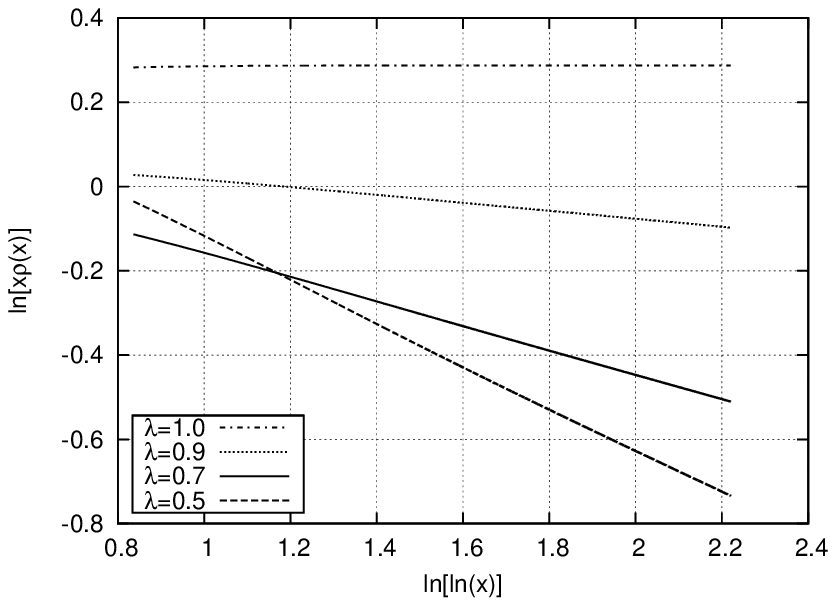}
\end{center}
\caption{Logarithmic dependence of the eigenvalue density  for $\lambda$-ensembles for $\lambda \le 1$.} 
\label{KNxx-fit01}
\end{figure}\begin{figure}[tbp]
\begin{center}
\includegraphics[angle=0, width=0.32\textheight]{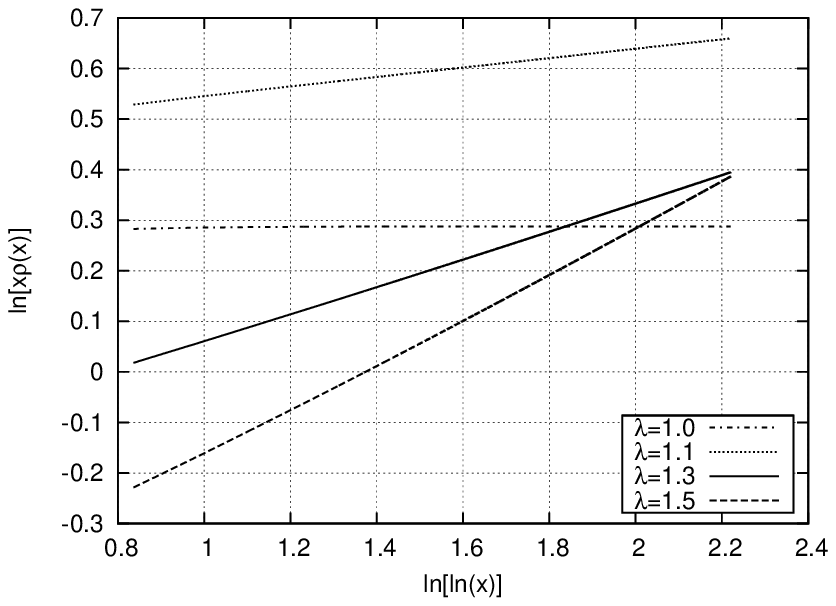}
\end{center}
\caption{Logarithmic dependence of the eigenvalue density for $\lambda$-ensembles for $\lambda \ge 1$.} 
\label{KNxx-fit02}
\end{figure}

In order to test the validity of Eq.~(\ref{eq-density}) numerically, 
we assume that the density is in general given by   
\begin{equation}
\rho(x) = \frac{f(x;\lambda)}{x+\Lambda}
\end{equation}
where $f(x;\lambda)$ is a slowly varying function of $x$. Then $x\rho (x)$ will have the form
\begin{eqnarray}
x\rho(x) &=& \frac{x}{x+\Lambda} f(x;\lambda), \\
              &\simeq&  f(x;\lambda)  \;\;\; x \gg \Lambda.
\end{eqnarray}
Thus, the function $f(x;\lambda)$ is identifiable in the large $x \gg \Lambda$ regime of $x\rho(x)~\textrm{vs}.~ x$ plot.
Figure \ref{KNxx-01} shows the logarithmic behavior of $x\rho(x)$ for $\lambda <1 $ $(\lambda = 0.5, 0.7, 0.9,  \textrm{and}~ 1)$, while Figure \ref{KNxx-03} show it for  $\lambda>1$ $(\lambda = 1.1, 1.3, 1.5, \textrm{and} ~1)$, respectively.  For all the cases, we chose $\gamma =  O(1) $, which ensures the cut-off $\Lambda = O(1)$. 


To further investigate if $f(x;\lambda) \propto [\ln x]^{\lambda -1} $ for large $x$ ($x \gg 1$), we plot $\ln[x\rho(x)]$ vs. $\ln \ln x$ and fit it in the range $10< x < 10^4 $. 
Figures \ref{KNxx-fit01} and \ref{KNxx-fit02} show the expected linear behavior, establishing the validity of Eq.~(\ref{eq-density}).

\subsection{Ghost correlation peak}

\begin{figure}[tbp]
\begin{center}
\includegraphics[angle=0, width=0.35\textheight]{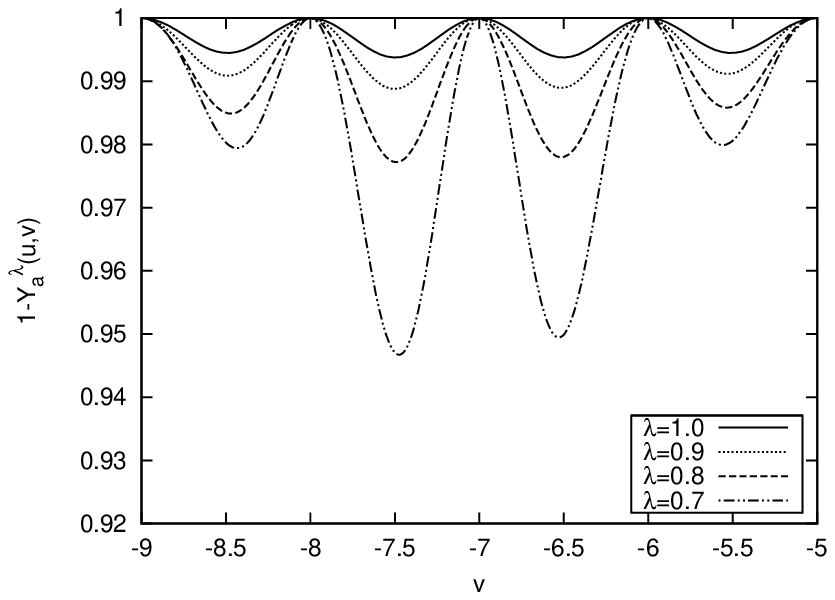}
\end{center}
\caption{Anomalous component of the cluster function for $\lambda$-ensembles for $\lambda  \le 1$ ($\gamma$ = 0.50).} 
\label{Yanom-01}
\end{figure}
\begin{figure}[tbp]
\begin{center}
\includegraphics[angle=0, width=0.35\textheight]{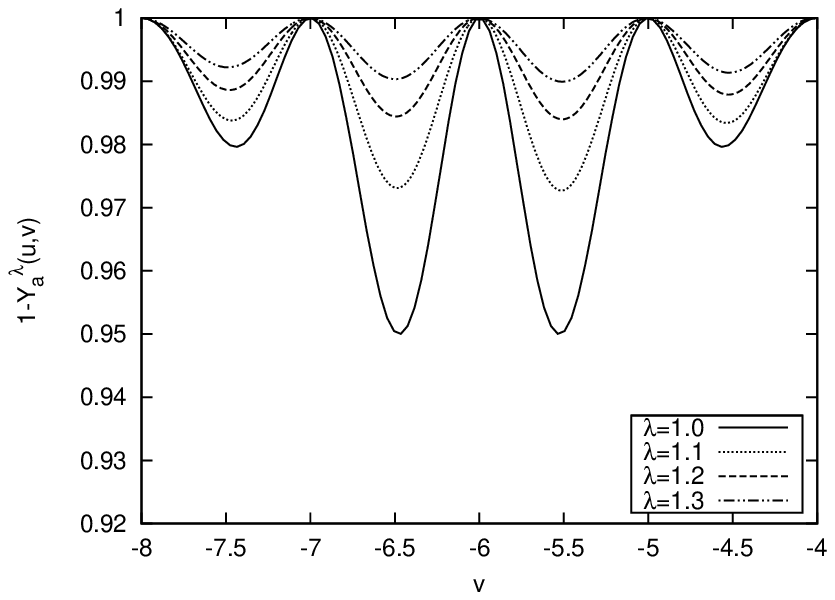}
\end{center}
\caption{Anomalous component of the cluster function for $\lambda$-ensembles for $\lambda \ge 1$ ($\gamma$ = 1.50).} 
\label{Yanom-02}
\end{figure}

For $\lambda$ =1, the existence of a ``ghost correlation peak'' is well known \cite{mean-field,qrme2,ghost-peak,sum-rule}. The ghost peak was first discussed in Ref. [\onlinecite{sum-rule}] though it was
contained implicitly in the exact result of Ref. [\onlinecite{qrme2}]. While the normal part of the two-level cluster function for the critical ensembles for $uv>0$ and $|u-v| << u$  is given, in the $\gamma \ll 2 \pi^2$ limit,  by the sinh kernel
\begin{equation}
Y_{n}^{c} (u, v) =  \Big[ \frac{\gamma }{2 \pi } \frac{  \sin [\pi (u - v)]}{ \sinh [  \frac{\gamma}{2} (u-v) ] } \Big]^{2},
\label{eq-Ynorm}
\end{equation}
there exists an
anomalous part of the cluster function for $uv <0$ given by
\begin{equation}
Y_{a}^{c} (u, v) =  \Big[ \frac{\gamma }{2 \pi } \frac{  \sin [\pi (u - v)]}{ \cosh [  \frac{\gamma}{2} (u+v) ] } \Big]^{2}.
\label{eq-Yanom}
\end{equation} 
In fact, the presence of such long range correlation is required by the normalization sum rule \cite{sum-rule}
\begin{equation}
1= \int_{-\infty}^{\infty} du~ [ Y_n^{c}(u,u') +Y_a^{c}(u,u')].
\label{sumrule}
\end{equation}
The deficiency of the sum rule 
\begin{equation}
\chi \equiv 1 - \int_{-\infty}^{\infty} du Y_n^{c}(u,u') =  \int_{-\infty}^{\infty} du Y_a^{c}(u,u')
\end{equation}
is related to certain characteristics of the critical statistics \cite{critical}: i) the level compressibility in the number variance $\Sigma(L)$ within a range $L$ and ii) the multi-fractality of eigenvectors. In particular,   
\begin{equation}\label{eq-multi}
\chi = \frac{d\Sigma(\langle L \rangle)}{d \langle L \rangle}  = \frac{d-D_2}{2d},
\end{equation}
where $d$ is the system dimension and the fractal dimensionality $D_p$ determines the scaling behavior of the moments of the inverse participation ratio via
\begin{equation}
\langle \int d^{d}x |\phi(x)|^{2p} \rangle \propto L^{-D_p(p-1)}.
\end{equation} 

It turns out that $\lambda$-ensembles possess such normal/anomalous structure for $\lambda \ne 1$ as well. Figures~\ref{Yanom-01} and \ref{Yanom-02} show the numerical evaluation of 1- $Y^{\lambda}_a(u,v)$ ($u>0$ and $v<0$) for a symmetric range around $v = -u$ for varying $\lambda$ values for a fixed $\gamma$ ($\gamma =0.5$ in Fig.~\ref{Yanom-01} and $\gamma=1.5$ in Fig.~\ref{Yanom-02}). As the figures show clearly, the magnitude of the ghost peak depends on $\lambda$ in a significant way; for $\lambda <1$ (Fig.~\ref{Yanom-01}), the peak is more pronounced than that for $\lambda=1$ and for $\lambda>1$ (Fig.~\ref{Yanom-02}) it is the opposite.

The observation that such long range correlation leading to the ghost peak is preserved for all $\lambda \neq 1$ suggests that such features are common to all logarithmic confinement potentials. In other words, once the critical ensembles break the $U(N)$ symmetry \cite{sum-rule} of the Gaussian ensembles with the introduction of the parameter $\gamma$ (or $q = e^{-\gamma}$), the $\lambda$ ensembles remain in this broken symmetry family. The fact that as $\lambda$ becomes large, the ghost peak shrinks seems to imply that the $U(N)$ symmetry might become fully restored in the limit of $\lambda \to \infty$. This expectation seems consistent with the asymptotic behavior of two-level correlation in the limit of $\lambda \to \infty$ that will be shown later in section D.

The $\lambda$-ensembles are all ``critical'' in the sense that the normal part of the two-level kernel violates the sum rule that can be associated with the characteristics of the critical statistics such as level compressibility and multi-fractality. In particular, the fact that the violation of the sum rule is controlled by the parameter $\lambda$, e.g., $0< \chi(\lambda)<1$ is intriguing. As mentioned above, $\lambda$ seems related to the degree of the $U(N)$ symmetry breaking and thus, indicative of the non-trivial character of the eigenvector correlations, namely the multi-fractal dimensionality as can be seen immediately from Eq.~(\ref{eq-multi}) and the $\lambda$-dependent sum-rule deficiency. In this regard, the study of the dimensional dependence of the critical statistics will be important to further understand the role of the parameter $\lambda$ since the multi-fractal dimension of the eigenvector correlations at the critical states is dependent on the dimensionality of the system \cite{D-dependence}.

\begin{figure}[tbp]
\begin{center}
\includegraphics[angle=0, width=0.35\textheight]{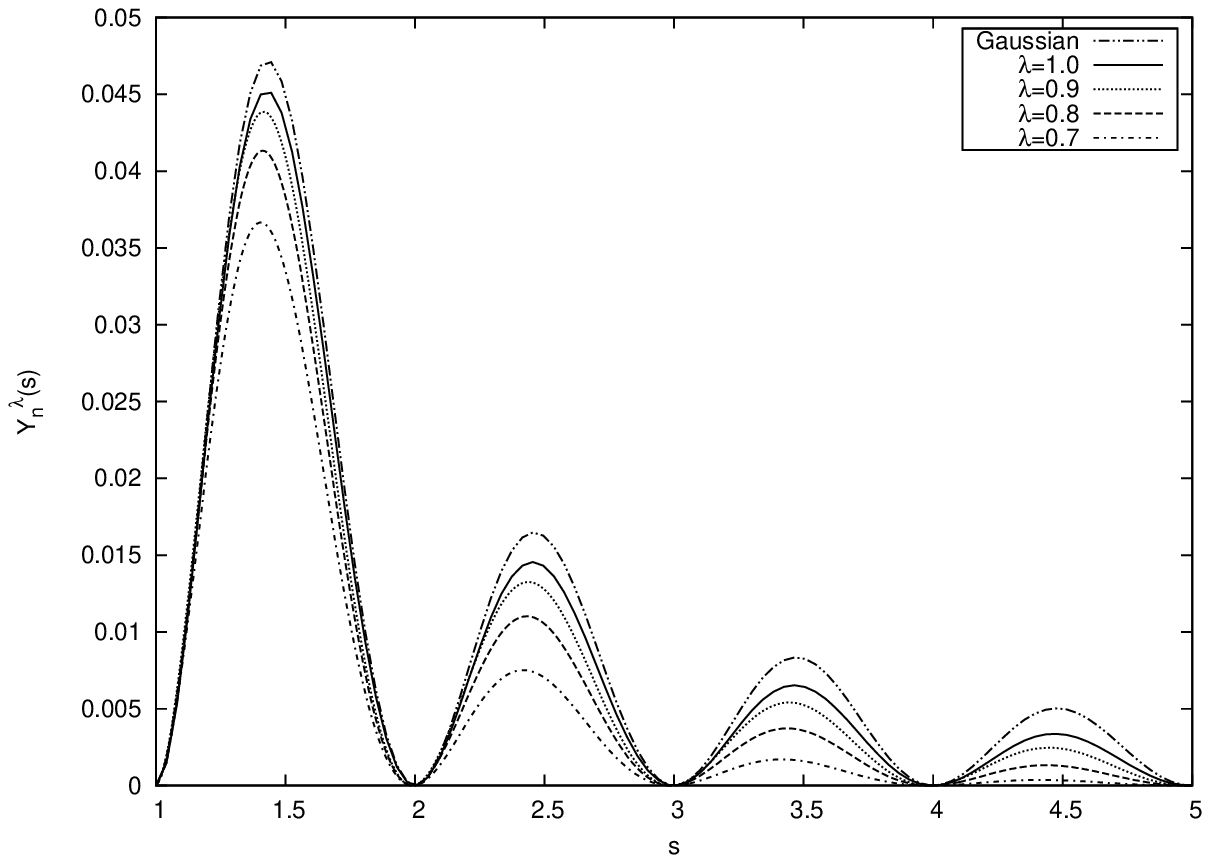}
\end{center}
\caption{Normal component of the cluster function for $\lambda$-ensembles for $\lambda \le 1$ ($\gamma$ = 0.50)} 
\label{Y-01}
\end{figure}
\begin{figure}[tbp]
\begin{center}
\includegraphics[angle=0, width=0.35\textheight]{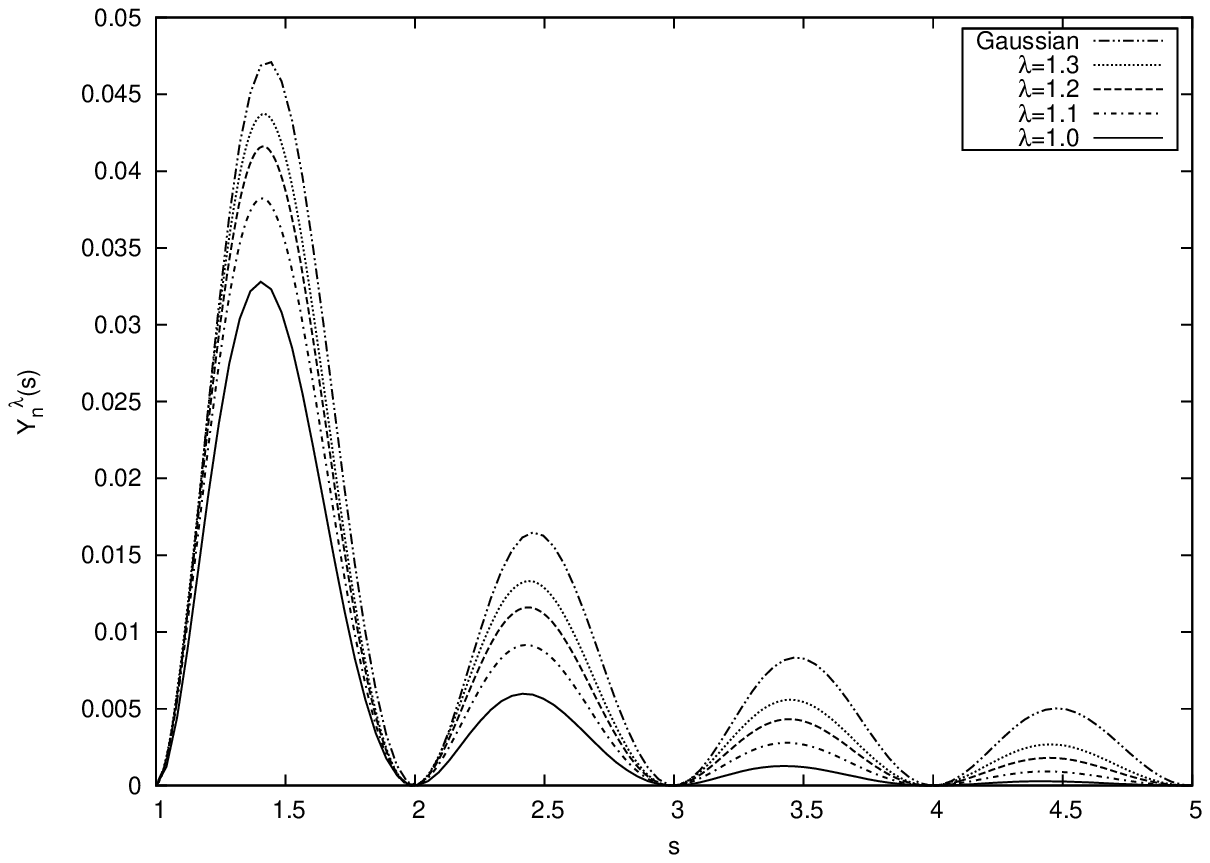}
\end{center}
\caption{Normal component of the cluster function for $\lambda$-ensembles for $\lambda \ge 1$ ($\gamma$ = 1.50)} 
\label{Y-02}
\end{figure}

Figures \ref{Y-01} and \ref{Y-02} show the comparison between the normal part of cluster function for $\lambda \neq 1$ and that for $\lambda = 1$ as well as that of the Gaussian ensemble.  We note that the nodes of the cluster functions occur at integer values on the $x$ axis for all $\lambda$ values as well as the Gaussian case, which is the indication of the semi-classical behavior. The peak height and position between the nodes gradually decrease and shifts as $\lambda$ decrease for a given $\gamma$ value ($\gamma =0.5$ in Fig.~\ref{Y-01} and $\gamma=1.5$ in Fig.~\ref{Y-02} ), which is already shown in Ref. [\onlinecite{cm-jpa09}].

\subsection{Identification of two-level kernel}

As mentioned before, for the logarithmic soft-confinment potential, the eigenvalue density is a non-trival function that do not depend on $N$ as shown in section A. The deformation from the Gaussian universality, the `sine kernel', is expected due to the non-trivial unfolding procedure \cite{unfolding,Kravtsov}. In the following, we will identify the two-level kernel of the $\lambda$-ensembles in the semi-classical regime and show that the proposed form is in good agreement with the numerically-obtained cluster function (thus the two-level kernel). 

In the semi-classical regime ($\gamma < 2\pi$ for the critical ensembles corresponding to $\lambda =1$), the kernel for an arbitrary weight function can be written as \cite{Kravtsov,semi-classical}
\begin{equation} 
\label{eq-semi-cls}
\bar{K}(u, v) \sim \frac{ \sin( \pi (u  - v)  )}{ x(u) - y(v) }.
\end{equation}  
For example, for the critical ensembles ($\lambda = 1$) the unfolding variable, in which the density becomes uniform and unity, is given by 
\begin{eqnarray}
u'=u-u_0  \equiv \int_{x_0}^{x} \frac{c}{t^{1}} dt =   \ln x/x_0.
\end{eqnarray} 
This leads to $x = x_0 e^{ \frac{u'}{c} }$ and hence to the $\sinh$ kernel. 
In a similar manner, we define the unfolding variable for the $\lambda$-ensembles as 
\begin{eqnarray}
u'= u-u_0 \equiv \int_{x_0}^{x} \frac{c [\ln t]^{\lambda-1}}{t} dt =  \frac{c}{\lambda} ( [\ln x]^{\lambda}-[\ln x_0]^{\lambda} ) 
\end{eqnarray}
where $u_0 = \int_0^{x_0} \rho(x) dx$ and $x_0 \gg \Lambda$, the cut off of the density near the origin. The constant $c \propto \rho(x=0)$. Note that $\rho(x=0)$ is a function of the parameters that do not depend on $N$.  For the critical ensembles, $\rho(0) = \frac{1}{\gamma}$ and for the $\lambda$-ensembles, $\rho(0) =g(\lambda,\gamma)$ where $g$ is some function of $\lambda$ and $\gamma$.
Rewriting the original variable $x$ in terms of the unfolding variable, we obtain 
\begin{equation}
x = \exp \left[  \gamma^{*} (u' + a )^{1/\lambda}  \right] 
\end{equation} 
where $ \gamma^{*} \equiv \left(\frac{\lambda}{c}\right)^{1/\lambda}$ and the $a \equiv \frac{c}{\lambda} [ \ln x_0]^\lambda$. 
For $\lambda = 1$, it reduces to the unfolding for critical ensembles.
Eq.~(\ref{eq-semi-cls}) then suggests the two-level kernel for $\lambda$-ensembles to be of the form
 \begin{equation}
{\bar K}_{n} (\tilde{u}, \tilde{v}) =  \frac{\Gamma }{2 \pi } \frac{  \sin [\pi(\tilde{u} -\tilde{v})]}{ \sinh [\frac{\Gamma}{2} ( \tilde{u} -\tilde{v})]},
\end{equation} 
where we have defined $\tilde{u} = u'+a $ and $\tilde{v} = v'+a$, and the parameter $\Gamma$ is introduced to satisfy the condition that $ \sigma(\tilde{u}) = {\bar K}_{n}^{\lambda} (\tilde{u},\tilde{u}) = 1$.
It then follows that 
\begin{equation}
\Gamma=\Gamma (\tilde{u}, \tilde{v}) \equiv \gamma^{*} \frac{\tilde{u}^{1/\lambda} -\tilde{v}^{1/\lambda} }{\tilde{u} - \tilde{v} }.
\end{equation} 
Thus we arrive at the regular component of the two-level kernel of the $\lambda$-ensembles
 \begin{equation}
 \label{eq-Kn}
{\bar K^\lambda}_{n} (\tilde{u}, \tilde{v}) =  \frac{\Gamma(\tilde{u},\tilde{v}) }{2 \pi } \frac{  \sin [\pi(\tilde{u} -\tilde{v})]}{ \sinh [\frac{\Gamma (\tilde{u},\tilde{v})}{2} (\tilde{u} - \tilde{v}) ]}; \;\;\; \tilde{u}\tilde{v} > 0,
\end{equation} 
while the anomalous component of the two-level kernel of the $\lambda$-ensembles is given by
 \begin{equation}\label{eq-K-anom}
{\bar K^\lambda}_{a} (\tilde{u}, \tilde{v}) =  \frac{\Gamma(\tilde{u},\tilde{v}) }{2 \pi } \frac{  \sin [\pi(\tilde{u} -\tilde{v})]}{ \cosh [\frac{\Gamma (\tilde{u},\tilde{v})}{2} (\tilde{u} + \tilde{v}) ]}; \;\;\; \tilde{u}\tilde{v} < 0 .
\end{equation}

We note that the kernel Eq.~(\ref{eq-Kn}) reduces to the well-known $\sinh$ kernel for $\lambda = 1$, where $ \Gamma(\tilde{u},\tilde{v}) \to \gamma^{*}=\gamma$. In general it is not translationally invariant, e.g. for $\lambda =0.5, \Gamma(\tilde{u},\tilde{v}) = \gamma^{*}({\tilde{u} +\tilde{v}})$ and for $\lambda=2.0, \Gamma(\tilde{u},\tilde{v}) = \frac{\gamma^{*}}{ \sqrt{\tilde{u}} +\sqrt{\tilde{v}}}$.
However, if we choose $\tilde{v}$ to be a fixed value, \textit{i.e}, $\tilde{v}= v-u_0 +a = a$, which is the same as choosing $v=u_0$,  then $\tilde{u}= u-u_0 + a = u-v+a \equiv s+a$. In this way, the function $\Gamma(\tilde{u}, \tilde{v})$ and $g(\tilde{u}, \tilde{v})$  can be written in terms of a difference variable $s \equiv u-v = \tilde{u} -\tilde{v}$ alone with a constant $\tilde{v}= a$ that serves as a fixed reference point,
\begin{equation}
\Gamma (s+a,a) \equiv \Gamma (s,a) = \gamma^{*} \frac{{(s+a)}^{1/\lambda} - {a}^{1/\lambda} }{s}.
\end{equation} 
Then 
\begin{equation}\label{Kn-sa}
{\bar K^\lambda}_{n} (s, a) =  \frac{\Gamma(s,a)}{2 \pi } \frac{  \sin (\pi s )}{ \sinh [\frac{\Gamma (s,a)}{2} s  ]}.
\end{equation}

\begin{figure}[h!]
\begin{center}
\includegraphics[angle=0, width=0.34\textheight]{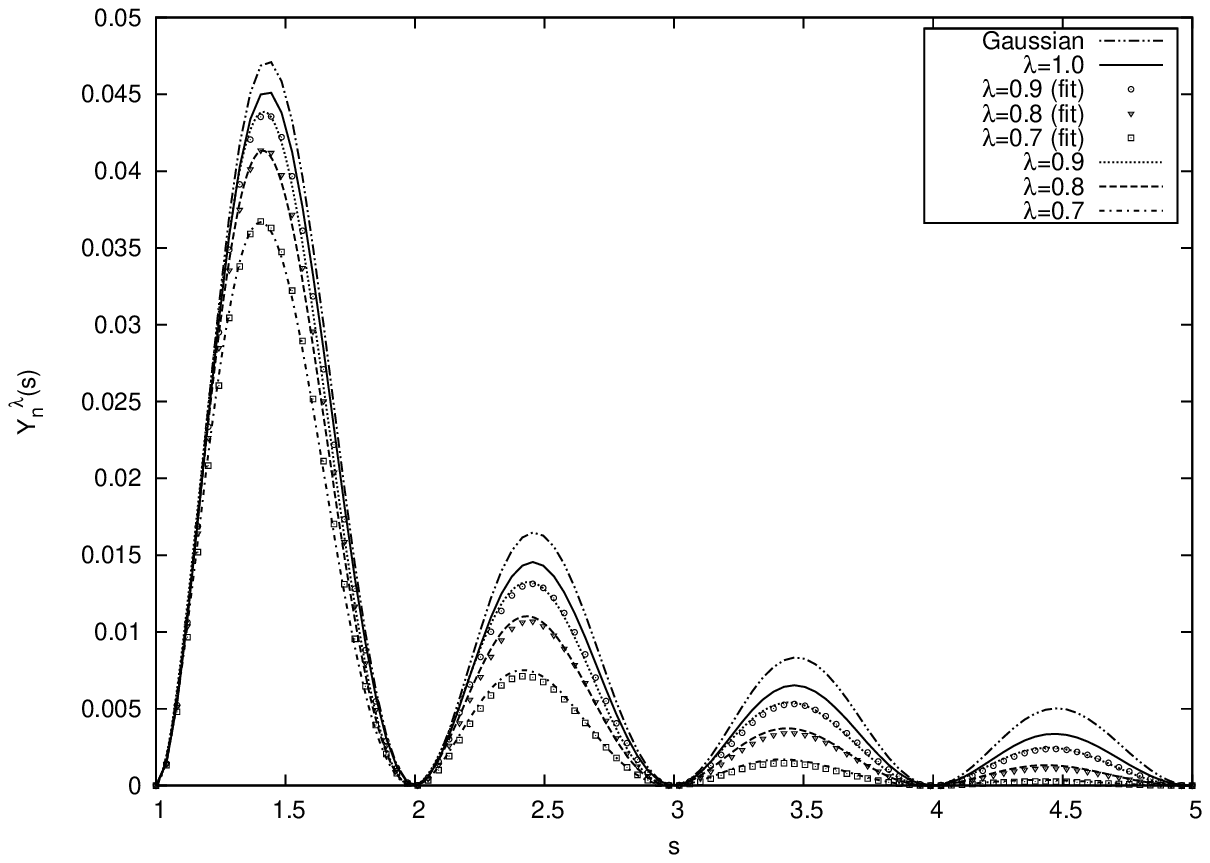}
\end{center}
\caption{Fitting results for normal component of the cluster function for $\lambda$-ensembles for $\lambda \le 1$  ($\gamma$ = 0.50). The lines are obtained from Eq.~(\ref{K-N}) using the $\lambda$-generalization of the $q$-polynomials, while the points are obtained from the analytic form of the cluster function given by Eq.~(\ref{eq-Kn}).} 
\label{Y-01-fit}
\end{figure}

\begin{figure}[tbp]
\begin{center}
\includegraphics[angle=0, width=0.34\textheight]{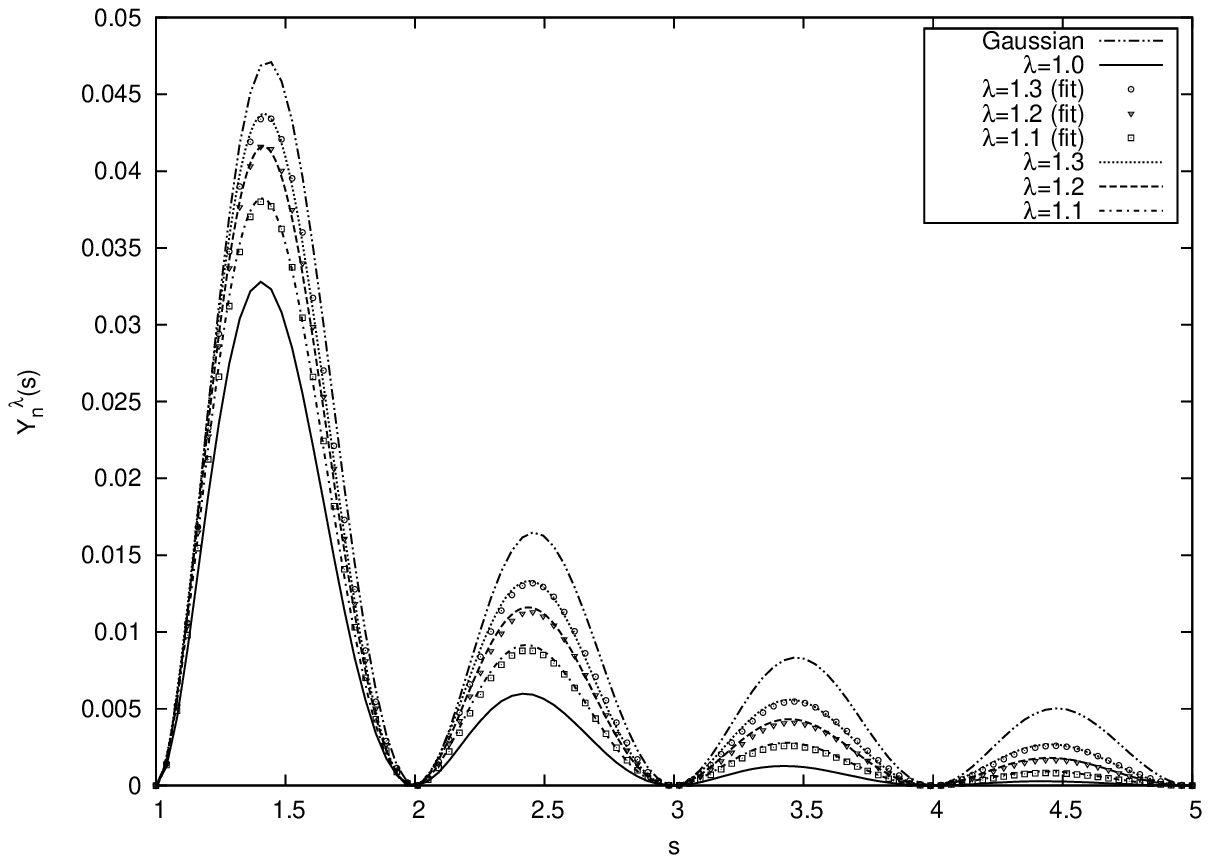}
\end{center}
\caption{Fitting results for normal component of the cluster function for $\lambda$-ensembles for $\lambda \ge 1$ ($\gamma$ = 1.50). The lines are obtained from Eq.~(\ref{K-N}) using the  $\lambda$-generalization of the $q$-polynomials, while the points are obtained from the analytic form of the cluster function given by Eq.~(\ref{eq-Kn}).} 
\label{Y-02-fit}
\end{figure}

\begin{figure}[tbp]
\begin{center}
\includegraphics[angle=0, width=0.35\textheight]{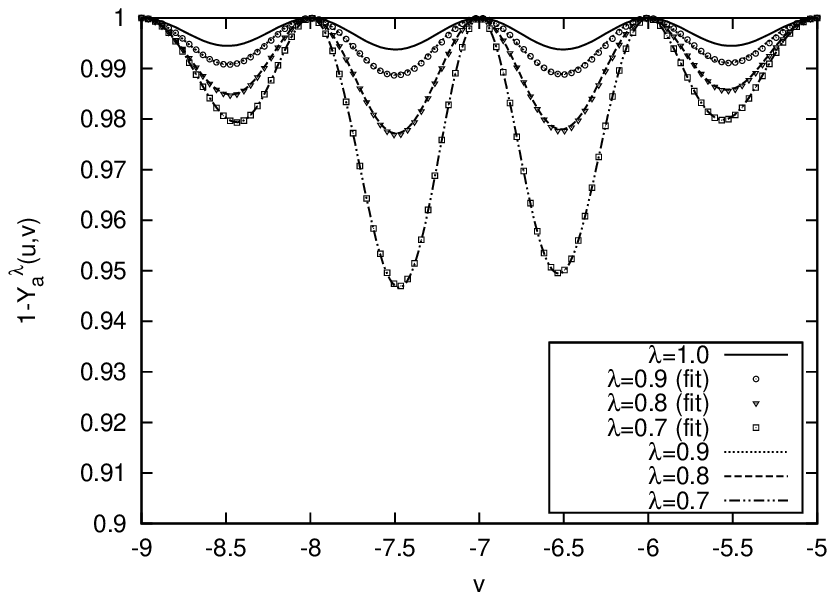}
\end{center}
\caption{Fitting results for anomalous component of the cluster function for $\lambda$-ensembles for $\lambda \le 1$ ($\gamma$ = 0.50).  The lines are obtained from Eq.~(\ref{K-N}) using the  $\lambda$-generalization of the $q$-polynomials, while the points are obtained from the analytic form of the cluster function of Eq.~(\ref{eq-K-anom}).} 
\label{Yanom-01-fit}
\end{figure}

\begin{figure}[tbp]
\begin{center}
\includegraphics[angle=0, width=0.35\textheight]{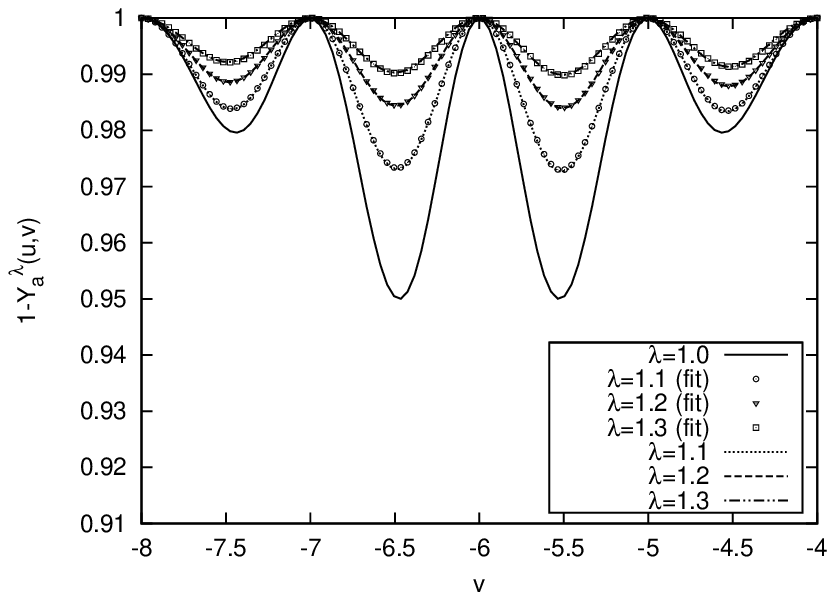}
\end{center}
\caption{Fitting results for anomalous component of the cluster function for $\lambda$-ensembles for $\lambda \ge1$ ($\gamma$ = 1.50). The lines are obtained from Eq.~(\ref{K-N}) using the  $\lambda$-generalization of the $q$-polynomials, while the points are obtained from the analytic form of the cluster function of Eq.~(\ref{eq-K-anom}).} 
\label{Yanom-02-fit}
\end{figure}

Figures \ref{Y-01-fit} and \ref{Y-02-fit} show the fitting results with the cluster function corresponding to the kernel Eq.~\ref{Kn-sa} for different $\lambda$ values. They show a fit with the numerically-obtained two-level kernel, from the $\lambda$-generalization of the $q$-polynomials and Eq.~(\ref{K-N}), with fit values $\gamma^{*} \simeq \gamma = 0.5$ and $a \simeq 2$ for $\lambda <1$ (Fig. \ref{Y-01-fit}) and  $\gamma^{*} \simeq \gamma = 1.5$ and $a \simeq 2$ for $\lambda >1$ (Fig. \ref{Y-02-fit}). 
In Figures \ref{Yanom-01-fit} and \ref{Yanom-02-fit}, we show that in similar range of the parameters, the anomalous components of the kernel also agree very well with the proposed form of the kernel Eq~(\ref{eq-K-anom}).

\subsection{Novel universality}
In the following, we will discuss novel asymptotic behavior of the two-level kernel of the $\lambda$-ensembles by further examining Eq.~(\ref{Kn-sa}). First, we note that 
\begin{eqnarray}
\Gamma(s,a) \propto s^{\frac{1}{\lambda} -1} \;\;\; \textrm{for} \;s \gg a 
\end{eqnarray}
leads to 
\begin{eqnarray}
\bar{K}_n(s,a) \propto s^{\frac{1}{\lambda} -1}e^{ -s^{\frac{1}{\lambda}}} \;\;\; \textrm{for} \;s \gg a .
\end{eqnarray}
Looking at the $\lambda$-dependence, we observe that 
\begin{eqnarray}
\label{K-asymptotic}
\bar{K}_n(s,a) &\propto& e^{-s^{\infty}}  \; \textrm{for} \; \lambda \to 0 \cr
&\propto& e^{ -s} \;\;\;\; \textrm{for} \;\; \lambda = 1 \cr 
&\propto& s^{-1} \;\;\;\; \textrm{for} \;\; \lambda \to \infty .
\end{eqnarray}
Thus, in the limit $\lambda \to \infty$, we get back the well-known decay of the Gaussian ensembles (WD universality). This is consistent with the fact that the magnitude of the ghost correlation peaks become smaller as $\lambda$ increases, presumably disappearing in the Gaussian limit of very large $\lambda$.  For $\lambda \to 0$ limit, the asymptotic tail is given by the infinitely fast exponential decay, which suppresses the eigenvalue correlations, thereby leading to uncorrelated Poisson-like behavior (Poisson statistics). In between these limits, the large $s$ behavior is governed by $e^{-s^{1/\lambda}}$, which is a novel feature of $\lambda$-ensembles. 

In a similar fashion, we can study the two-level density-density correlation function  
\begin{equation}
R(u,v) \equiv \delta(u-v) - Y(u,v)
\end{equation}
In terms of the difference variable $s = u-v$, the two-level correlation function for the $\lambda$-ensembles can be written as 
\begin{equation}
R^{\lambda}(s) \equiv \delta(s) - Y^{\lambda}(s) 
\end{equation}
For $s \ll a$,  $R^{\lambda}(s)  \sim s^2$  for all values of $\lambda$, which is an expected feature due to the unitary symmetry of the ensembles. For large $s \gg a$, the results are simply obtained from Eq.~(\ref{K-asymptotic})
\begin{eqnarray}\label{eq-novle-R}
R^{\lambda}(s,a)&\propto& e^{-s^{\infty}} \; \textrm{for} \; \lambda \to 0 \cr
&\propto& e^{ -s} \;\;\; \textrm{for} \;\; \lambda = 1 \cr
&\propto& s^{-2} \;\;\; \textrm{for} \;\; \lambda \to \infty 
\end{eqnarray}
We can see that the asymptotic behavior of the two-level correlation function interpolates that of Gaussian ensembles for $\lambda \to \infty$ limit and uncorrelated Poisson-like behavior for $\lambda \to 0$ limit.  This suggests that $\lambda$-ensembles belong to the critical ensembles characterized by $\lambda$ which determines the rate of asymptotic exponential decay of the two-level kernel.

\subsection{Number variance}

\begin{figure}[tbp]
\begin{center}
\includegraphics[angle=0, width=0.35\textheight]{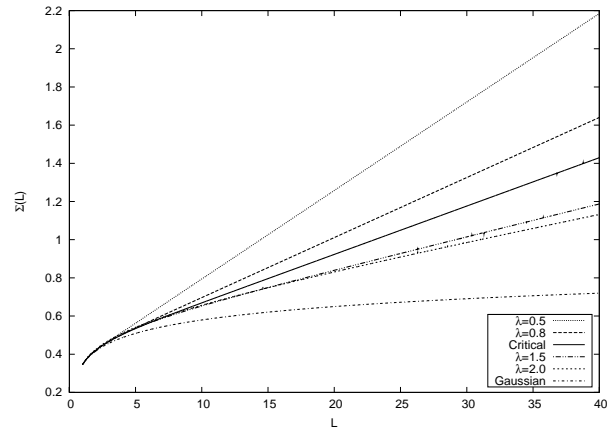}
\end{center}
\caption{Number variance for $\lambda$-ensembles for $\lambda =$0.5, 0.8, 1.0, 1.5, 2.0 ($\gamma$ = 0.50).} 
\label{nv}
\end{figure}

As an example of the spectral measures obtained from the two-level kernel, the evaluation of the number variance $\Sigma(L)$ within a range $L$ is shown in Figure~\ref{nv}. It clearly shows that for all values of $\lambda$, the number variance is linear in $L$ for large $L$.  As $\lambda$ becomes smaller, the number variance shifts towards the uncorrelated Poisson distribution. This is consistent with the fact that  deficiency of sum rule rule $\chi(\lambda)$ increases as $\lambda$ decreases as shown in section B, since the slope of number variance is directly related to the deficiency of the sum rule:
\begin{equation}
\chi (\lambda)= \frac{d\Sigma(\langle L \rangle)}{d \langle L \rangle}.
\end{equation}

\section{Summary and Discussion}

In this work, we further study an invariant-class of random matrix ensembles characterized by the asymptotic logarithmic soft-confinement potentials introduced in Ref.~[\onlinecite{cm-jpa09}] which we refer to as the $\lambda$-ensembles. As a first step, we carefully reinvestigate the spectral density of the $\lambda$-ensembles and show that instead of a $\lambda$-dependent power law in a restricted regime, the spectral density is given more accurately by a power-law of the form $\rho(x) \propto [\ln x]^{\lambda-1}/x$. This result is suggested by the mean-filed approach and it can be checked by the normalization condition of the spectral density, the edge of which is determined by the coefficient $R_n$ of the $\lambda$-generalization of the $q$-polynomials. 

Second, we show that the two-level kernel of the $\lambda$-ensembles has normal/anomalous structure, which is characteristic of the critical ensembles. The anomalous component arising due to the sum rule violation is dependent on the parameter $\lambda$ for a given value of $\gamma$ which is kept $O(1)$ in the numerics; As the value of the $\lambda$ decreases, the deficiency of the sum rule becomes larger.

Third, we identify the normal and anomalous components of the two-level kernel in the semi-classical regime, which are given by Eq.~(\ref{eq-Kn}) and Eq.~(\ref{eq-K-anom}) that reduce to those of the critical ensembles for $\lambda=1$. Further, we show that the two-level kernel of the $\lambda$-ensembles exhibits a distinct universal asymptotic behavior, shown by Eqs.~(\ref{K-asymptotic}), which includes the Gaussian ensembles ($\lambda \to \infty$ limit), the critical ensembles ($\lambda = 1$) as well as the uncorrelated Poisson-like behavior ($\lambda \to 0$ limit). In particular, the large $s$ behavior of the two-level kernel is governed by $\exp[-s^{1/\lambda}] $, which is the characteristic feature of the $\lambda$-ensembles. It is expected that the asymptotic tail of the spacing distribution is also given by a similar exponential form.

Lastly, we show that the number variance is linear in $L$ for large $L$. The slope of number variance or the level compressibility is dependent on the parameter $\lambda$. As $\lambda$ decreases, the slope increases, which is consistent with the fact that anomaly or the deficiency of the sum rule increases as $\lambda$ decreases.

All the features of the $\lambda$-ensembles shown above such as the ghost peak, the deficiency of the sum-rule, the finite compressibility as well as the asymptotic exponential decay of the kernel seem to suggest  that the $\lambda$-ensembles belong to the critical ensembles characterized by $\lambda$. Therefore, it is expected that the $\lambda$-ensembles would be relevant to the description of the critical states of the localization-delocalization problems in disordered systems. Since the critical level statistics are universal, depending only on the critical exponent and the dimensionality of system for a given symmetry class, it is conceivable that the parameter $ \lambda$ can be associated with these parameters. In particular, the study of the dimensional dependence of the critical statistics is interesting since the parameters of the critical level statistics such as the critical exponent and the multi-fractal dimension are dependent on the spatial dimension of the system.  

At the same time, it is also interesting to see if $\lambda$-ensembles are also applicable in the study of quantum chaos. Recently the critical statistics has been found relevant in some cases of quantum chaos as well \cite{Garcia}. It turns out that the two level kernel of chaotic systems with logarithmic singularity \cite{Garcia-Wang} have the exact same form as that of the critical ensembles. 

Another important implication of the $\lambda$-ensembles is in regard to the question that we posed earlier, namely if there is a universality of the correlations of the eigenvalues associated with fat-tail distributions. Our results seem to suggest that a non-trivial $N$-dependence of two-level kernel of the fat-tail RMEs is a generic feature within the framework of rotationally-invariant RMT. In the $\lambda \to 0$ limit of the $\lambda$-ensembles, $\gamma$ is required to be an $N$-dependent parameter to have the probability measure to be normalizable, which is the case for the free L\'evy matrices. Thus, the $N$-dependence of the two-level kernel in this limit can be understood as a consequence of the presence of the $N$-dependent parameter in the model ensemble, which cannot be simply scaled out. 

In conclusion, we show that the family of RMEs with the asymptotic logarithmic soft-confinement potentials characterized by $\lambda$, called the ``$\lambda$-ensembles", connects the WD universality ($\lambda \to \infty$), the uncorrelated Poisson-like behavior ($\lambda \to 0$) and exhibit a critical behavior for all the intermediate $\lambda$ value ($0<\lambda<\infty $) in the semi-classical regime. We expect that further study of the $\lambda$-ensembles and the $\lambda$-generalization of $q$-polynomials will lead to a deeper understanding of the universality of random matrix ensembles in general.

We gratefilly acknowledge valuable discussions with V.E. Kravtsov.

\end{document}